\documentclass[aps,pra,twocolumn]{revtex4}

\usepackage{makeidx}
\usepackage{amsfonts}
\usepackage{amssymb}
\usepackage{enumerate}
\usepackage{latexsym}
\usepackage{bbm}
\usepackage{graphicx}
\usepackage{graphics}
\usepackage{here}

\newcommand{\beq}{\begin{equation}}
\newcommand{\eeq}{\end{equation}}
\newcommand{\beqa}{\begin{eqnarray}}
\newcommand{\eeqa}{\end{eqnarray}}

\newcommand{\identity}{\hat{\openone}}

\newcommand{\sumgren}[2]{\sum\limits_{#1}^{#2}}

\newcommand{\ket}[1]{\left\vert #1 \right\rangle}
\newcommand{\bra}[1]{\left\langle #1 \right\vert}
\newcommand{\bracket}[3]{\bra{#1} \hat{\sigma}_{#2} \ket{#3}}

\begin{document}

\title{Entanglement quantification through local observable correlations}

\author{Christian Kothe}
\affiliation{School of Information and Communication Technology,
Royal Institute of Technology (KTH), Electrum 229, SE-164 40 Kista,
Sweden}
\author{Gunnar Bj\"{o}rk}
\affiliation{School of Information and Communication Technology,
Royal Institute of Technology (KTH), Electrum 229, SE-164 40 Kista,
Sweden}

\date{\today}

\begin{abstract}

We present a significantly improved scheme of entanglement detection
inspired by local uncertainty relations for a system consisting of
two qubits. Developing the underlying idea of local uncertainty
relations, namely correlations, we demonstrate that it's possible to
define a measure which is invariant under local unitary
transformations and which is based only on local measurements. It is
quite simple to implement experimentally and it allows entanglement
quantification in a certain range for mixed states and exactly for
pure states, without first obtaining full knowledge (e.g. through
tomography) of the state.

\end{abstract}

\maketitle

\section{Introduction}

Entanglement is one of the key resources in quantum mechanics and in
particular in quantum communication and quantum information
applications. Its appearance and behavior was discovered and
discussed a long time ago \cite{EPR,SR}, but it was only since the
1980s that applications of entanglement, like quantum teleportation
\cite{Ben}, quantum cryptography \cite{BB84,Eke} and quantum
algorithms \cite{Deu,Sho,Gro}, were developed and became the focus
for intense research.

An important question which is not answered in full generality yet
is: Assuming a given state, how much ``profit'' (in terms of
entanglement) is inherently hidden in that state that could be used
to perform some of the tasks mentioned above? A state with higher
entanglement should allow us to perform a task in some sense better
than a state with lower entanglement. Therefore a lot of effort has
been spent during the last two decades to investigate entanglement
further, and in particular, to quantify it \cite{HHH,Hor}, so that
states can be quantitatively ranked. For the case of two qubits,
substantial progress has been achieved and entanglement of formation
and concurrence are widely accepted as well behaved and
operationally meaningful measures for entanglement \cite{Woot2}.
However, for higher-dimensional cases and multipartite states the
situation gets more complicated and, despite some progress, the
search for good measures is still going on \cite{RB,AF,EG}.

Even if good measures exist for the case of two qubits, they
require, in general, full knowledge of the density matrix for a
given state to be determined. This is achieved by full state
tomography \cite{Du}, a cumbersome and time-consuming experimental
measurement process. A way to avoid these inconveniences is to use
so-called entanglement witnesses \cite{LK,LK2,GL}, which can detect
specific entanglement, but, on the other hand, are not able to
quantify it. An alternative to entanglement witnesses are local
uncertainty relations (LUR) \cite{HT}. They are typically easy to
implement experimentally, but unfortunately no known LUR can detect
all entangled two-qubit states, and in general they do not give
quantitative measure of entanglement for the states they do detect.
Assuming that one has an unknown state, it is therefore desirable to
quantify its entanglement as well as possible, with the lowest
possible experimental effort.

In this  paper we will extend the idea of local uncertainty
relations for two qubits and thereby overcome most of its drawbacks,
but keeping it's advantages. We will derive a measure that is
invariant under local unitary transformations, and which quantifies
entanglement for all pure states and in some range for mixed states.
An advantage is that the measure only requires local measurements
(in contrast to \cite{BH}, for example), facilitating the
experimental effort. A mathematically similar approach as ours has
been taken by de Vicente \cite{Vicente}, who use the Bloch-vector
representation of bipartite states and the Ky Fan norm to write an
inequality that can only be broken by nonseparable states. However,
Vicente does not address how his entanglement criteria can be
experimentally implemented.

Compared to state tomography our measure will require the same
number of measurement-settings for the case of two qubits (9
settings), but we conjecture significantly less measurement-setups
in higher-dimensional systems. On the other hand one can always
calculate the concurrence after a state tomography, since one has
full knowledge of the state. The simplest LUR requires only 2
measurement-settings \cite{KH} but LURs are (except some special
cases) not able to quantify entanglement.

In the next section we will motivate the measure starting from local
uncertainty relations and then discuss it's properties. After that
we will investigate the case of pure states in section \ref{sec pure
states} and the case of mixed state in section \ref{sec mixed
states}, before summarizing the results and discuss still open
questions in section \ref{sec summary}.

\section{Definition and implementation}

Before giving the definition of the new measure, we will motivate it
by giving a short review of entanglement detection through local
uncertainty relations. Even though the theory of local uncertainty
relations has been extended to multipartite systems by G\"{u}hne
\cite{OG}, we will only cover bipartite systems here. Having two
systems $A$ and $B$, one can choose sets of observables
$\{\hat{A}_i\}$ and $\{\hat{B}_i\}$, acting solely on the
corresponding system. In each set of observables it is assumed that
the observables have no joint eigenvector. The local variances are
then given by
$\delta^2\hat{A}_i\equiv\left\langle\hat{A}_i^2\right\rangle-\left\langle\hat{A}_i\right\rangle^2$,
and similar for $\delta^2\hat{B}_i$. These variances are
nonnegative, and for the variance $\delta^2\hat{A}_i$ to be zero the
system has to be in an eigenstate of $\hat{A}_i$. Because the
operators $\{\hat{A}_i\}$ have no joint eigenstates the variance
$\delta^2\hat{A}_j$ for this state must be positive for all $j \neq
i$. Therefore, there exist a state (or a class of states) that have
an associated non-trivial value $U_A>0$ which is the greatest lower
limit of the sum of the variances. From this follows that for any
state the following inequality holds: \beq
\sumgren{i}{}\delta^2\hat{A}_i\geq U_A. \eeq The same thing holds
for the observables of system $B$ and we will define $U_B$ in the
same way, so that \beq \sumgren{i}{}\delta^2\hat{B}_i\geq U_B. \eeq
The operators $\hat{A}_i+\hat{B}_i$ can be defined to measure
properties of the common system. One can show that the local
uncertainty relation \beq \label{deflur}
\sumgren{i}{}\delta^2\left(\hat{A}_i+\hat{B}_i\right)\geq U_A+U_B
\label{eq: inequality} \eeq then holds for all statistical mixtures
of product states. A proof for (\ref{eq: inequality}) for this class
of states is given in \cite{HT}.

Expanding the left hand side of Eq. (\ref{deflur}) gives \beqa
\nonumber \sumgren{i}{}\delta^2\left(\hat{A}_i+\hat{B}_i\right) & =
& \sumgren{i}{}\delta^2\hat{A}_i+\sumgren{i}{}\delta^2\hat{B}_i
\\ && +2\sumgren{i}{}C\left(\hat{A}_i,\hat{B}_i\right), \eeqa where the
covariance term is defined as \beq \label{defcorr}
C\left(\hat{A}_i,\hat{B}_i\right)=\left\langle\hat{A}_i\hat{B}_i\right\rangle-\left\langle\hat{A}_i\right\rangle\left\langle\hat{B}_i\right\rangle.
\eeq To reveal entanglement by not fulfilling inequality
(\ref{deflur}), one can immediately see, that at least one of the
covariance terms has to be less than zero for such an entangled
state. Any single covariance term is bounded by \beq
-\left(\delta^2\hat{A}_i+\delta^2\hat{B}_i\right)\leq
2C\left(\hat{A}_i,\hat{B}_i\right)\leq\delta^2\hat{A}_i+\delta^2\hat{B}_i.
\eeq Since both bounds can be reached with both mixed separable and
pure entangled states for any particular choice of a pair of
observables $\hat{A}_i$ and $\hat{B}_i$, one has to look at several
covariances to detect entanglement. To give an example, one can
consider a two-level system, e.g., the polarization states of
spatially separated photon pairs. A possible LUR in this case is
\beqa \nonumber
L_3 & = & \delta^2\left(\hat{\sigma}^A+\hat{\sigma}^B\right)_{0/90}+\delta^2\left(\hat{\sigma}^A+\hat{\sigma}^B\right)_{45/135}\\
& & +\delta^2\left(\hat{\sigma}^A+\hat{\sigma}^B\right)_{R/L}\geq 4,
\label{eq: Howell} \eeqa where the subscript 0/90 denotes
measurements of horizontal and vertical linear polarization. Assume
that the measurement eigenvalues for $\hat{\sigma}^A$ and
$\hat{\sigma}^B$ are $\pm 1$. This means that the possible
measurement outcomes for $\hat{\sigma}^A+\hat{\sigma}^B$ are -2, 0,
and 2. The subscript 45/135 denotes similar measurements in a basis
rotated by 45 degrees and R/L denotes measurements of left- and
right-handed polarized photons. The relation (\ref{eq: Howell}) was
investigated by Ali Khan and Howell in \cite{KH} and $L_3\geq4$ is
fulfilled by all mixtures of separable states, but may be violated
for entangled states, the minimum of $L_3$ being zero in this latter
case. The lower bound $L_3=0$ is attained by the singlet state
$(\bra{\uparrow,\downarrow} - \bra{\downarrow,\uparrow})/\sqrt{2}$
because this state has perfectly anticorrelated polarization if
photon $A$ and $B$ are measured in any same basis. Therefore $L_3$
can detect entanglement. Note, however, that {\em $L_3$ assumes a
shared spatial reference frame,} because if $A$ and $B$ are measured
with the respective horizontal and vertical axes unaligned, $L_3$
will no longer be zero for the singlet state. In addition, only a
small fraction of the set of entangled qubit states is detected by
$L_3$ and a local unitary transformation of a given entangled state
is sufficient to make a violated LUR fulfilled, or vice versa,
although the entanglement remains invariant per definition. As an
example, a local basis-state flip on either $A$ and $B$ (which is
equivalent to an interchange of the vertical and the horizontal
axis) results in the state $(\bra{\uparrow,\uparrow} -
\bra{\downarrow,\downarrow})/\sqrt{2}$ for which $L_3$ takes its
maximum value eight, well over the threshold for entanglement
detection, which is four. This example demonstrates the necessity of
a shared spatial reference frame. If we, on the other hand, were
using a measure which is invariant under local unitary
transformations we would not have to align our measurement setups,
since every local rotation can be described by a local unitary
transformation. Especially if the measurement devices are located
far apart this can lower the experimental effort significantly.

An attempt to rectify some of the problems with LURs was done in
\cite{SB}, where an improved way of using local uncertainty
relations, so called modified local uncertainty relations (MLUR),
was proposed. These MLUR could detect more states than LUR, but the
main drawbacks of local uncertainty relations, namely invariance
under local unitary transformations, remained. An advantage of LURs,
compared to state tomography or entanglement witnesses, is the
relatively small experimental effort which is needed to implement
them experimentally. Therefore we propose in this paper a new
measure inspired by local uncertainty relations, which keeps the
advantages of LUR like low experimental effort, but gets rid of some
of the disadvantages of LUR, for example being not invariant under
local unitary transformations, not detecting all entangled pure
states or not quantifying entanglement.

One realizes that the information of entanglement is somehow coded
in the covariances defined by Eq. (\ref{defcorr}), since only they
are responsible for violating a LUR. We propose therefore for the
case of two qubits to use the sum of all possible covariances
between two local sets of mutually unbiased bases, one for each
qubit, \beq \label{defG} G=\sumgren{i,j=1}{3}
C^2\left(\hat{\sigma}_i^A,\hat{\sigma}_j^B\right) , \eeq as a
measure of entanglement. Here, $\hat{\sigma}_i^A$ denotes the $i$:th
Pauli matrix (operator) for system $A$, and similar for $B$. The
Pauli matrices are \beq \label{Pauli}
\hat{\sigma}_1=\left(\begin{array}[c]{cc}0 & 1\\
1 & 0
\end{array}\right),
\hat{\sigma}_2=\left(\begin{array}[c]{cc}0 & -i\\
i & 0
\end{array}\right),
\hat{\sigma}_3=\left(\begin{array}[c]{cc}1 & 0\\
0 & -1
\end{array}\right),
\eeq whose eigenvalues are $\pm 1$. The Pauli matrices are
traceless: $\mathrm{Tr}\left(\hat{\sigma}_i\right)=0$ (for
$i=1,2,3$). In the following we will denote the unity matrix as
$\hat{\sigma}_0$: \beq
\hat{\sigma}_0=\left(\begin{array}[c]{cc}1 & 0\\
0 & 1
\end{array}\right). \label{eq: identity}
\eeq  These four matrices have the property
$\mathrm{Tr}\left(\hat{\sigma}_i \hat{\sigma}_j\right)=2\delta_{ij}$
for $i,j=0, \ldots,4$. The measure of Eq. (\ref{defG}) is easy to
implement. Since \beq
C\left(\hat{\sigma}_i^A,\hat{\sigma}_j^B\right)=\left<\hat{\sigma}_i^A\otimes\hat{\sigma}_j^B\right>-\left<\hat{\sigma}_i^A\otimes\identity^B\right>\left<\identity^A\otimes\hat{\sigma}_j^B\right>,
\eeq one has only to count singles rates and coincidences. The
measurements can be performed locally on each system and a total of
nine measurement-settings are sufficient
($\left<\hat{\sigma}_i\otimes\identity\right>$ and
$\left<\identity\otimes\hat{\sigma}_j\right>$ can be calculated from
the other measurements) to get all the results. Here, we would like
to point out that $G$ is invariant under any local unitary
transformations. That is,
$G\left(\hat{\rho}\right)=G\left(\hat{U}\hat{\rho}
\hat{U}^\dagger\right)$ with $\hat{U}=\hat{U}_A\otimes \hat{U}_B$
where $\hat{U}_A$ ($\hat{U}_B$) operates only on subsystem A (B).
($G$ is also invariant under partial transposition.) The invariance
is a direct consequence of the fact that our measure $G$ can be
rewritten as \beq \label{HilSch}
G=4\mathrm{Tr}\left\{\left(\hat{\rho}-\hat{\rho}_A\otimes\hat{\rho}_B\right)^2\right\},
\eeq that is, in terms of the Hilbert-Schmidt norm measure of
distance, where $\hat{\rho}_A$ denotes the density matrix of system
$A$ after tracing over system $B$ and similar for $\hat{\rho}_B$
\cite{Hall}. Since any unitary transformation can be seen as a
rotation or mirroring of the basis vectors of a Hilbert-space and a
norm is invariant under rotation or mirroring the basis, it follows
that $G$ is invariant under any local unitary transformation. To see
the equivalence between Eq. (\ref{defG}) and Eq. (\ref{HilSch}) one
can expand the density matrix $\hat{\rho}_A$ as \beq \label{rhoA}
\hat{\rho}_A=\frac{1}{2}
\sumgren{n=0}{3}\mathrm{Tr}\left(\identity\otimes\hat{\sigma}_n\hat{\rho}\right)\hat{\sigma}_n
\eeq and similar for $\hat{\rho}_B$. $\hat{\rho}$ can also be
expanded in the basis defined by the operators $\hat{\sigma}_i$ as
\beq \label{rho}
\hat{\rho}=\frac{1}{4}\sumgren{n=0}{3}\sumgren{m=0}{3}\mathrm{Tr}\left(\hat{\sigma}^A_m\otimes\hat{\sigma}^B_n\hat{\rho}\right)\hat{\sigma}^A_m\otimes\hat{\sigma}^B_n.
\eeq Inserting the expansion (\ref{rhoA}) and its $\hat{\rho}_B$
counterpart, and (\ref{rho}) into (\ref{HilSch}), and using the
Pauli matrix relations written just under (\ref{Pauli}) and
(\ref{eq: identity}), it is not difficult to rewrite the ensuing
equation in the form (\ref{defG}).

{\em Note that, because of the local unitary invariance is it not
necessary to use the Pauli matrices in the definition of our
proposed measure in Eq. (\ref{defG}). Every local unitary
transformation of this mutually unbiased basis (MUB) \cite{KW} works
equally well.} The invariance under local unitary transformations
also means that {\em a shared spatial reference frame is no longer
needed}, because a local rotation (a unitary transformation) will
leave the measure invariant. However, since it is sufficient to use
the Pauli matrices and since they are convenient from an
experimental and mathematical viewpoint, we will continue to use
them in this paper.

For pure states $G$ is just a bijective function of the
well-established concurrence, whereas for mixed states $G$ relates a
state to a certain range of concurrence. The proof of these
statements are given next.

\section{Pure states}
\label{sec pure states}

We will relate our measure to the well known concurrence in the case
of pure states. For the proof we will expand an ordinary pure
two-qubit state into the eigenvectors $\ket{00}$, $\ket{01}$,
$\ket{10}$ and $\ket{11}$ of the
$\hat{\sigma}_3^A\otimes\hat{\sigma}_3^B$ operator, that is \beqa
\nonumber
\ket{\psi} & = & a_{00}\ket{00}+a_{01}\ket{01}+a_{10}\ket{10}+a_{11}\ket{11}\\
& = & \sumgren{k,l=0}{1}\alpha_{kl}\ket{k}^A\otimes\ket{l}^B.
\eeqa
The adjoint state can be written in a similar way and, dropping the sign for the tensor product, we can expand
\beqa
\nonumber
G & = & \sumgren{i,j=1}{3}\left(\left<\hat{\sigma}_i^A\hat{\sigma}_j^B\right>^2-2\left<\hat{\sigma}_i^A\hat{\sigma}_j^B\right>\left<\hat{\sigma}_i^A\right>\left<\hat{\sigma}_j^B\right>\right.\\
\label{Ggrund} & &+ \left.\left<\hat{\sigma}_i^A\right>^2
\left<\hat{\sigma}_j^B\right>^2\right) \eeqa in a sum of products of
the expansion coefficients $\alpha_{kl}$.

Recall now the results of Linden and Popescu \cite{LP}, where they
derive the invariants of systems of different dimensions under local
unitary transformations. For the case of two qubits they show that
there are only two invariants: \beqa
I_{1} & = & \sumgren{k,l=0}{1}\alpha_{kl}\alpha_{kl}^{\ast}\\
I_{2} & = &
\sumgren{k,l,m,n=0}{1}\alpha_{km}\alpha_{kn}^{\ast}\alpha_{ln}\alpha_{lm}^{\ast}.
\eeqa

Evaluating the expansion of Eq. (\ref{Ggrund}) performing
considerable trivial, but tedious, algebra (i.e. evaluating all the
terms $\bracket{\cdot}{}{\cdot}$ and making the summation), the
result is \beq \label{Gfertig}
G=\left(I_{\alpha}^{2}+8I_{\beta}\right)-2I_{\alpha}\left(I_{\alpha}^{2}-4I_{\beta}\right)+\left(I_{\alpha}^{2}-4I_{\beta}\right)^2,
\label{eq: first G}\eeq where the first term in Eq. (\ref{Gfertig})
corresponds to the first term in the Eq. (\ref{Ggrund}) and so on.
The summation over $i$ and $j$ is already included in each term.
$I_{\alpha}$ and $I_{\beta}$ stand for \beqa
I_{\alpha} & = & \left|\alpha_{00}\right|^2+\left|\alpha_{01}\right|^2+\left|\alpha_{10}\right|^2+\left|\alpha_{11}\right|^2\\
I_{\beta} & = &
\left(\alpha_{01}\alpha_{10}-\alpha_{00}\alpha_{11}\right)\left(\alpha_{01}^{\ast}\alpha_{10}^{\ast}-\alpha_{00}^{\ast}\alpha_{11}^{\ast}\right)\\
&=& \left | \alpha_{00}\alpha_{11} - \alpha_{01}\alpha_{10} \right
|^2 . \eeqa One sees immediately that $I_{\alpha}=I_{1}$.  If one
looks at Eq. (\ref{Gfertig}) in further detail one sees that
$I_{\alpha}$ is not just a constant under unitary transformations,
it is simply the state normalization constant and therefore
$I_{\alpha}=1$ for all states. One can, after some algebra, also
show that $I_{\beta}=(I_{1}^{2}-I_{2})/2$. Therefore $I_{\beta}$ is
also an invariant. This simplifies equation (\ref{eq: first G}) to
\beq G=8I_{\beta}+16I_{\beta}^2=4I_{\beta}\left(2+4I_{\beta}\right).
\eeq Knowing that $4I_{\beta}={\cal C}^2$ \cite{AF}, where ${\cal
C}$ denotes the well-known concurrence \cite{Woot2} for pure states,
we finally get \beq \label{GzuC} G={\cal C}^2\left(2+{\cal
C}^2\right). \eeq The concurrence is related to entanglement of
formation and having therefore a pure state, one can directly
quantify its entanglement by measuring $G$. For a pure state $G>0$
implies that the state is entangled. This is an intuitive result
because a separable, pure state cannot display any covariance
between local measurements.

\section{Mixed states}
\label{sec mixed states}

\begin{center}
\begin{figure}
\center
\includegraphics[angle=0, scale=.55]{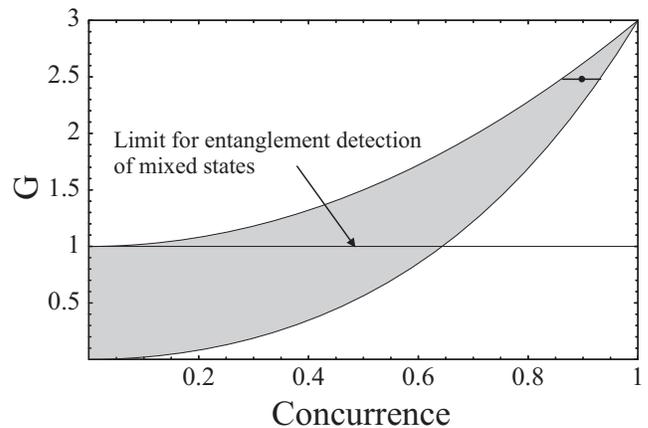}
\caption{Bounds for $G$ when plotted over the concurrence. See text
for more details.} \label{bounds}
\end{figure}
\end{center}

The relation between $G$ and the concurrence that held for the pure
states in Eq. (\ref{GzuC}) is no longer valid for mixed states.
Instead, $G$ can, in general, take any value in the shaded region
plotted in Fig. \ref{bounds}. That is, $0 \leq G \leq 3$ or, if we
write it in relation to the concurrence, one has \beq
\label{Ggrenzen} {\cal C}^2\left(2+{\cal C}^2\right)\leq G\leq 1+2
{\cal C}^2. \eeq The lower bound of this inequality is given by Eq.
(\ref{GzuC}), that is, any pure state has the lowest possible values
of $G$ for a given amount of entanglement. The reason for that is
that correlations for pure states can only be given by entanglement.
(Note that the converse is not true. That is, there are mixed states
that also saturate the lower bound of $G$.)

To find the upper bound of Eq. (\ref{Ggrenzen}) we look at the
density matrix \beq
\hat{\rho}_u=\left(\begin{array}[c]{cccc}1/2 & 0 & 0 & e^{i\theta}\gamma\\
0 & 0 & 0 & 0\\
0 & 0 & 0 & 0\\
e^{-i\theta}\gamma & 0 & 0 & 1/2
\end{array}\right)
\eeq with $0 \leq \gamma \leq 1/2$ and $\theta$ an arbitrary real
number. The class of states defined by this density matrix
interpolates between maximally classically correlated states with no
entanglement ($G=1$, ${\cal C}=0$ when $\gamma = 0$) and maximally
entangled (pure) states which have the highest correlations of any
states ($G=3$, ${\cal C}=1$ when $\gamma = 1/2$). It is reasonable
to believe that this class of states has the highest value of $G$
for any given amount of entanglement. Using the definition of $G$ in
Eq. (\ref{defG}) one finds that for these states
$G(\hat{\rho}_u)=1+8\gamma^2$. Calculating the concurrence gives
${\cal C}(\hat{\rho}_u)=2\gamma$ and therefore we have
$G(\hat{\rho}_u)=1+2{\cal C}^2$. A simulation with many thousands of
arbitrary states shows that, indeed, no state is outside the range
given by (\ref{Ggrenzen}). {\em Hence, a value of $G>1$ guarantees
that the system is entangled,} since $G$ of separable states (having
zero concurrence) cannot exceed unity.

Unfortunately, we don't have any strict algebraical proof for the
limits of $G$ at the moment although we firmly believe, and can see
clear arguments why the limits are both sufficient and necessary.
The problem is that it is not known how to parameterize the entire
class of states with a given concurrence, let alone to find the
maximum G of such a multiparameter class of states.

In general, $G$ is an ``entanglement witness'' for mixed states,
since it can detect entanglement for a class of states. But if one
has a state with a high concurrence, $G$ can give more information.
An example is given in Fig. \ref{bounds}. Imagine one measures a
value of $G=2.5$. In that case, the state has to have a concurrence
somewhere in the range of the upper horizontal line in Fig.
\ref{bounds}, that is \beq \label{Concbereich} 0.87 \leq {\cal
C}\leq  0.93. \eeq This is quite a narrow range. Hence, even if one
is not able to determine the concurrence exactly, $G$ is still able
to limit a state to a certain range of the concurrence. Assigning a
value of ${\cal C}=0.9$ in the case above would, for example, only
give a maximum error of $\pm4\%$ for the concurrence. $G$ is
therefore giving more information about a state than an entanglement
witness ordinarily does.

\begin{center}
\begin{figure}
\center
\includegraphics[scale=.55]{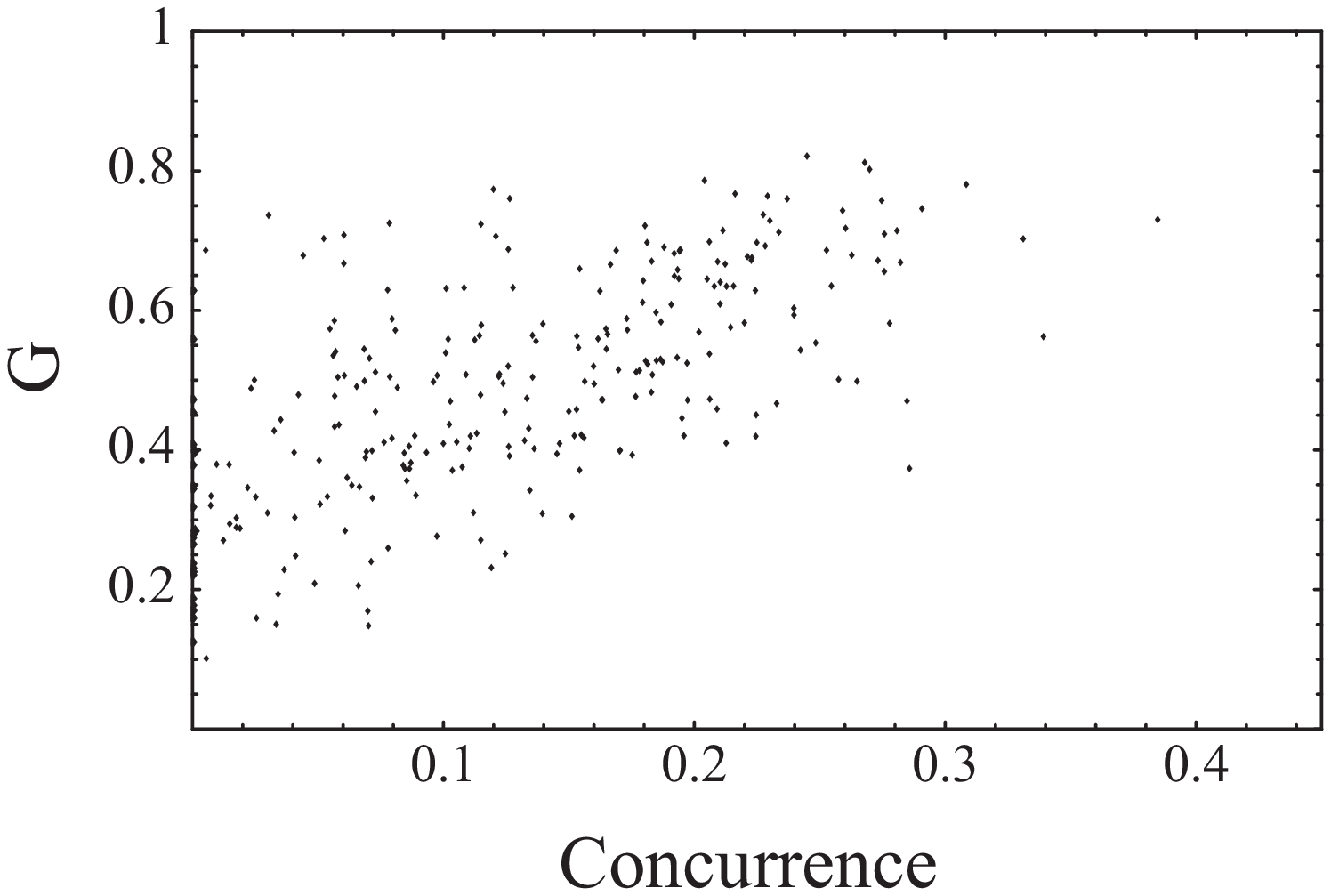}
\includegraphics[scale=.55]{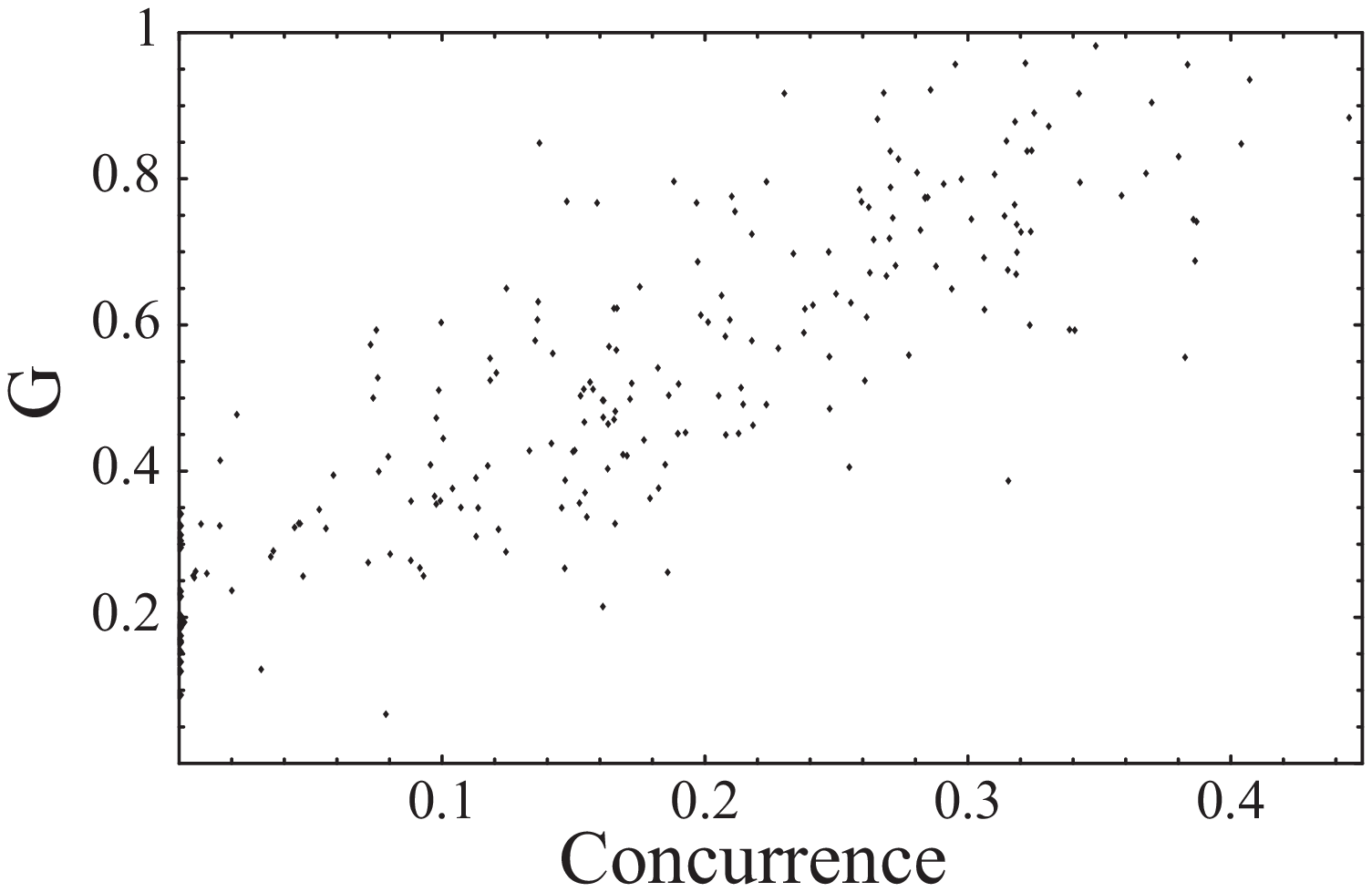}
\caption{Simulation results of measuring $G$ for some mixed states.
The upper figure is for states with a purity of
$\mathrm{Tr}\left(\hat{\rho}^2\right)=0.46\pm0.005$ and the lower
figure is for states with
$\mathrm{Tr}\left(\hat{\rho}^2\right)=0.5\pm0.005$.}
\label{mixedpic}
\end{figure}
\end{center}

An interesting question is, whether $G$ can somehow be
``compensated'' by the amount of mixedness, so that $G$ and ${\cal
C}$ become a bijective map also for mixed states. We have made some
simulations with arbitrary density matrices and plotted $G$ as a
function of the concurrence and of the degree of purity, defined by
$Tr\left(\hat{\rho}^2\right)$ (see \cite{MJ}). The result can be
seen in Fig. \ref{mixedpic}. If one fixes the concurrence to a
certain value and just looks at $G$ depending on the purity, it
turns out that these simulation results cover an area and not only a
line, showing that it is impossible, to write the measure as
$G=G\left({\cal C}\left(\hat{\rho}\right),
Tr\left(\hat{\rho}^2\right)\right)$ and thereby ``compensating'' $G$
by the purity. However, one might use other definitions than
$Tr\left(\hat{\rho}^2\right)$ for the mixedness to fulfill this.
This is still an open question.

\section{Summary and discussion}
\label{sec summary} Inspired by local uncertainty relations we have
suggested a measure of entanglement for two qubits, which can
quantify entanglement for pure states and can give bounds on the
entanglement of mixed states. This measure is invariant under local
unitary transformations and requires only local measurements to be
implemented. It might even be possible to get an exact
quantification for mixed states.

Further work will be focused on which properties a generalization of
our proposal would have for higher dimensional systems. In this
connection we can refer to Wootters \cite{Woot}, who showed that one
can determine all properties of a state by measuring all
combinations of local MUB eigenstate projections and the identity
matrix. We therefore conjecture that a generalization of our
proposed measure for higher-dimensional systems would keep the
properties like invariance under local unitary transformations and
can be useful to detect and quantify entanglement.

In that context the number of measurement-settings for our measure
would scale substantially lower as a function of the dimensions as
the number of measurement setups would scale for state tomography.
For multi-partite systems a similar method may still work, but in
this case the added complication that different kinds of
entanglement exist makes the problem both a quantitative
\textit{and} a qualitative one.

\section{Acknowledgements}

The authors want to thank Drs. Michael Hall and Otfried G\"{u}hne
for fruitful correspondence. This work was supported by the Swedish
Research Council (VR), the Swedish Foundation for Strategic Research
(SSF), and the European Community through grant Qubit Applications
\#015848.


\begin{thebibliography}{20}
\bibitem{Hall} We want to thank Dr. Michael Hall for pointing out this relation, and Ref. [21], to us.
\end{thebibliography}
\end{document}